\documentclass[11pt,twoside]{article}

\usepackage{asp2004}
\usepackage{epsf}
\usepackage{psfig}
\usepackage{lscape}
\def\bbe{\rm B$\rightleftharpoons$Be\ }

\begin{document}
\title{Evolution of the circumstellar disc of $\alpha$~Eri}
\author{Vinicius, M.M.F.$^1$, Leister, N.V.$^1$, Zorec, J.$^2$, Levenhagen 
R.S.$^1$}
\affil{$^1$Instituto de Astronomia, Geof\'{\i}sica e Ci\^encias Atmosf\'ericas 
da Universidade de S\~ao Paulo, Brazil}
\affil{$^2$Institut d'Astrophysique de Paris, UMR7095 CNRS, Univ. P\&M Curie}

\begin{abstract}
The H$\alpha$ line emission formation region in the circumstellar disc of 
$\alpha$~Eri is: a) extended with a steep outward matter density decline 
during low H$\alpha$ emission phases; b) less extended with rather constant 
density distribution during the strong H$\alpha$ emission. The long-term 
variation of the H$\alpha$ emission has a 14-15 year cyclic \bbe phase 
transition. The disc formation time scales agree with the viscous decretion 
model. The time required for the disc dissipation is longer than expected 
from the viscous disc model.
\end{abstract}
\vspace{-0.5cm}

\section{Characteristics of the H$\alpha$ line emitting region from 1991 to 
2002}
\begin{figure}[t]
\centerline{\psfig{file=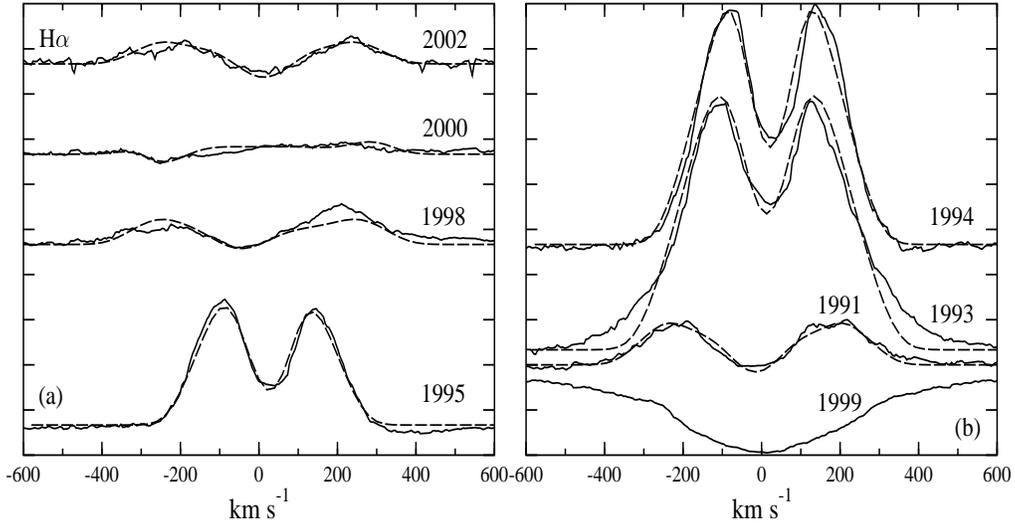,width=13.5truecm,height=6.8truecm}}
\caption[]{H$\alpha$ line profiles from 1991 to 2002. These profiles have the
photospheric line component (1999) subtracted, so that the continuum level is
set to zero. Observed profiles are in full lines and model fits are in dashed
lines. The scale of intensities $I/I_o$ is 0.15}
\label{f1}
\end{figure}

 The H$\alpha$ line of $\alpha$~Eri was observed several times from 1991 to 
2002, in particular, halfway through the epoch of the interferometric 
observations reported by Domiciano de Souza et al. (2003). The 1999 line 
profile is used to represent the photospheric absorption and subtracted from 
all observed H$\alpha$ emission lines. The `neat' H$\alpha$ line emission 
components are shown in Fig.~\ref{f1}.\par
 The source function $S$ of the H$\alpha$ line in $\alpha$~Eri is dominated by 
radiative ionization and recombination processes of atomic levels so that its 
dependence with the optical depth is (Mihalas 1978): $S_{H\alpha}(T_{\rm 
eff},\tau)\!=$ $\eta^{1/2}B^*$ for $\tau\!\leq\!1$, $\eta^{1/2}B^*\tau^{1/2}$
for $\tau\!>\!1$, where $\tau_o$ is the optical depth in the central 
wavelength of the line; $\eta$ is the radiative `sink' term; $B^*$ is the 
`source' factor dependent on photoionization and recombination rates. Using 
the ring-model described in Arias et al. (2005, this issue) and Vinicius et al.
(2005) to represent the H$\alpha$ emission line formation we derive the 
parameters given in Table~\ref{t1}: $\tau_o$, $R_r/R_o$ (ring radius); $H/R_*$ 
(semi-height of the ring); $V^r_{\Omega}$ (average ring rotation velocity); 
$V_{\rm rad}$ (average ring expansion velocity); $\beta$ form the particle 
density disitribution $N\sim R^{-\beta}$. The fits of the H$\alpha$ line 
emission component are also shown in Fig.~\ref{f1}. In all cases, a non 
negligible ring/disc effective height $H/R_*\!\sim\!3.4\pm0.6$ is required. 
The line profiles in 1991, 1998 and 2002 require on average quite extended 
emitting regions $<\!\!R_{\rm E}/R_o\!\!>\!\simeq\!40$ and outward steeply 
decreasing density distributions: $\beta\sim2$. The emitting region of 
H$\alpha$ from 1993 to 1995 is $<\!\!R_{\rm E}/R_o\!\!>\!\simeq\!11\pm3$ with
nearly the same disc height $<\!\!H/R_*\!\!>\!\simeq\!3.6\pm0.1$ and a quite 
uniform density distribution: $\beta\sim0$.

\begin{table}[]
\centering
\caption[]{Circumstellar disc parameters from fits of H$\alpha$ emission line 
profiles}
\label{t1}
\begin{tabular}{cccccrr}
\noalign{\smallskip}
\hline
\noalign{\smallskip}
Epoch  & $\tau_o$ & $R_r/R_o$  & $H/R_*$ & $V^r_{\Omega}$ & $V_{\rm rad}$
& $\beta$ \\ 
\noalign{\smallskip}
\hline
\noalign{\smallskip}
1991 & 0.15 & 3.8 & 3.8 & 265 &   0 &  1.8 \\ 
1993 & 1.18 & 6.0 & 3.5 & 197 &   0 &  0.2 \\ 
1994 & 0.70 & 5.9 & 3.8 & 185 &   0 & -0.1 \\ 
1995 & 0.25 & 5.5 & 3.5 & 190 & -10 &  0.1 \\
1998 & 0.09 & 3.8 & 3.5 & 275 &  50 &  2.0 \\
2000 & 0.08 & 3.8 & 2.2 & 220 & 255 & -0.4 \\
2002 & 0.13 & 3.7 & 2.5 & 270 & -15 &  1.9 \\
\noalign {\smallskip}
\hline
\end{tabular}
\end{table}

\section {Cyclic long-term H$\alpha$ line emission changes}

  We can take advantage of observations that we obtained on the H$\alpha$ line 
emission changes from 1991 to 2002, to derive some information on the time 
scales that characterize a complete cycle of \bbe transition. We also compare
this cycle with the long-term H$\alpha$ variations in $\alpha$~Eri observed in 
previous epochs. The time scales of these cycles matter to understand the disc 
formation mechanisms (Porter 1999, Okazaki 2001). We collected in the 
literature the spectroscopic and photometric records of the H$\alpha$ line 
variation and studied them as a function of time.\par
\begin{figure}[t]
\centerline{\psfig{file=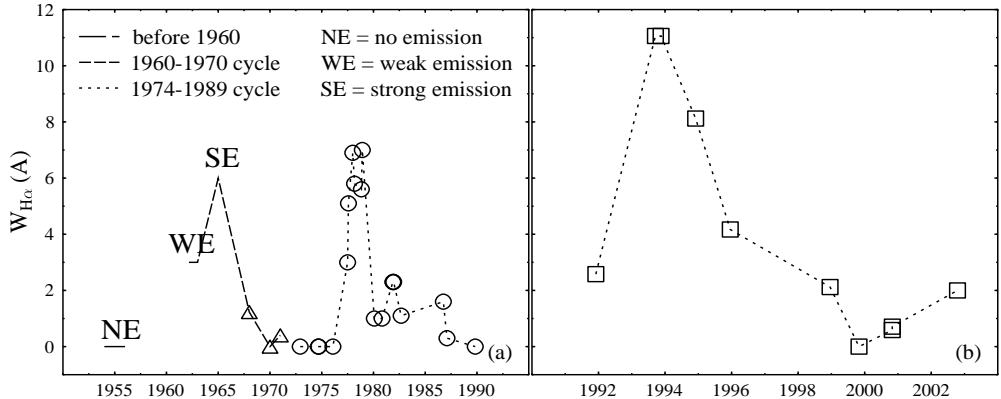,width=13.5truecm,height=5.5truecm}}
\caption[]{Long-term variation of the equivalent width $W$ in \AA\ of the 
H$\alpha$\ line emission. (a) Qualitative estimates of the emission strength 
(NE = no emission; WE =  weak emission; SE = strong emission) and quantitative 
equivalent widths before 1990 collected in the literature. (b) New equivalent 
widths obtained in this work after 1990}
\label{f2}
\end{figure}

  The variation of the equivalent width of the $H\alpha$ line emission 
component before 1991 is shown in Fig.~\ref{f2}a. In this figure we also 
added the qualitative estimates of the emission intensity noted in the 
literature before 1965. The changes of the $H\alpha$ emission observed from 
1991 to 2002 is shown in Fig.~\ref{f2}b. From both panels in Fig.~\ref{f2} we 
conclude that there is a sort of cyclic variation of the the $H\alpha$ line 
emission. The maxima of emission are attained each 14 years roughly, while the 
onsets of the increase towards the greatest emission maxima produce about each 
15 years. Fig.~\ref{f2} shows that the two last B-normal like aspects lasted 4
to 5 years. From the last two cycles we learn that after the quiescent, or 
B-normal like phases, the star resumes its maximum emission in no more than 2 
years. We can also see that after each emission maxima there is a 4-5 year 
lasting plateau of weak emission before the star recovers the B-normal like 
aspect. These time scales may have some significance to model the CE formation 
in this star.\par
 If the evolution of the CE in $\alpha$~Eri, i.e. formation and subsequent 
dissipation, has to be understood in terms of a viscous decretion disc model 
(Porter 1999, Okazaki 2001), from the expected average viscous time scales for
$t\sim60/\alpha$ days (Clark et al. 2003), we would derive two quite different 
series of values for the viscosity coefficient $\alpha$. The time scales 
implied by the CE formation, estimated from the time that takes the emission 
rise in the H$\alpha$ line: 2 years roughly, we would obtain $\alpha \sim$ 
0.08, which is of the same order as expected for CE envelopes studied by 
Blondin \& Negueruela (2001), Matsumo (1999) and Clark et al. (2003). The time
required by the decrease from maximum to zero is either 10 years in the 
1974-1989 cycle or 6 years during the 1991-2000 cycle. This imply $0.02
\la\alpha\la0.03$, which is about 4 times smaller than implied in the 
formation process. We may then wonder whether the bulk redistribution of 
material in the circumstellar disc occurs only as a consequence of a viscous 
redistribution of angular momentum.

\end{document}